%% file: opendave_nobugs.tex
\documentclass[12pt]{article}
\usepackage{graphicx}

\newcommand\pubnumber{NOBUGS2002/018}
\newcommand\pubdate{\today}

\def\frm2{ZBE FRM-II\\
Technische Universit\"at M\"unchen\\
Lichtenbergstr. 1\\
D-85748 Garching, F.R.G.\\
jens.krueger@frm2.tu-muenchen.de\\
}

\def\Title#1{\begin{center} {\Large #1 } \end{center}}
\def\Author#1{\begin{center}{ \sc #1} \end{center}}
\def\Address#1{\begin{center}{ #1} \end{center}}

\newcommand\pubblock{\rightline{\begin{tabular}{l} \pubnumber\\
         \pubdate  \end{tabular}}}
\newenvironment{Abstract}{\begin{quotation}  }{\end{quotation}}
\newenvironment{Presented}{\begin{quotation} \begin{center} 
             PRESENTED AT\end{center}\bigskip 
      \begin{center}\begin{large}}{\end{large}\end{center} \end{quotation}}

\input econfmacros.tex

\begin{document}
\begin{titlepage}
\pubblock

\vfill
\Title{\textbf{openDaVE} - a data visualisation and transformation framework}
\vfill
\Author{ Jens Kr\"uger}
\Address{\frm2}
\vfill
\begin{Abstract}
At present each neutron scattering facility has its own set of data treatment software, 
normally based on a proprietary data format. The travelling scientist is either forced 
to continually learn new software or to write conversion routines for his own software tools.

During software development most of the time is spent on routines connected with file 
input and output or visualisation. Our aim is to relieve the scientist from such tedious
work to allow maximum effort to be spent on data processing routines.

I present an  "Open Source" component based, extendable, platform independent framework 
for visualisation and transformation of data. 

Most of the existing software appears as a monolithic block. In contrast  \textbf{openDaVE} 
is built up in a modular design. It consists of three parts: the kernel, a lot of modules, and
some different front-ends. The actual functionality of the framework is done by the modules. The 
kernel provides the creation, the destruction of the modules, and the interaction between them, 
whereas the front-end interacts with the user.

The modules are divided into three classes: sources, filters, and sinks. Each of the module consists 
of two different parts: the module logic and one or more interfaces which set the module parameters. 
The interfaces are front-end specific. 

\end{Abstract}
\vfill
\begin{Presented}
NOBUGS 2002\\
Gaithersburg, U.S.A,  November 04--06, 2002
\end{Presented}
\vfill
\end{titlepage}
\def\thefootnote{\fnsymbol{footnote}}
\setcounter{footnote}{0}

\section{Introduction}
A lot of scientists in the neutron scattering community is travelling from one neutron source to another.
Most of them spent only some days at a certain instrument. 

During their work at different experiments at the various neutron sources they have to learn continually 
the use of the specific software packages for the pre-analysis of the data. To overcome this constant learning process 
they may use their own software, but they have to write new modules for accessing the data in the experiment 
specific data format.

For analysis and interpretation they get the experimental data in electronic form (on disc, by email, 
direct access, etc) mostly in an instrument specific data format, whereas the own software 
at home uses a significantly different data format. 

That means for a large amount of time a scientist is occupied by learning the handling of new software or 
by writing routines for data conversions. This problems lead to a project in the FRM-II software group to give 
all users a tool solving this problems: 
\begin{center}
\textbf{openDaVE} (\textbf{open} \textbf{Da}ta \textbf{V}isualization \textbf{E}nvironment).
\end{center}

\section{Features}

The main goal during the development was to achieve flexibility in many fields, which is provided by the following
design rules:
\begin{itemize}
\item component based architecture
\item simple extensibility
\item replaceable front-ends
\item replaceable data types
\end{itemize}

The \textbf{openDaVE} framework was written in C++ using the Qt library \cite{qt} under Linux. This decision allows 
a simple porting to other operating systems such as Windows and the different flavors of UNIX. The current version 
runs under Linux and Windows. 

\textbf{openDaVE} supports a lot of data formats, but one of the most interesting data formats is NeXus \cite{nexus}. 
The NeXus support includes: 
\begin{itemize}
\item reading and writing of NeXus file
\item browsing of the NeXus structure in a file 
\item conversion into NeXus format
\item export into non-NeXus formats
\item selection of data from a complex NeXus file containing more than one NXdata entry
\end{itemize}

In contrast to most of the existing software, \textbf{openDaVE} was not designed as a monolithic program. It
was built in a modular fashion, consisting of three parts:
\begin{itemize}
\item the front-ends
\item the modules
\item the kernel
\end{itemize}

This modularity is one most powerful features of \textbf{openDaVE}. It neither restricts the data used inside and outside 
\textbf{openDaVE} nor restricts anybody in the selection of functionality and user interface.

The different set-up's of \textbf{openDaVE} are saved in XML files. 

\section{System Design}

\begin{figure}[htc]
\begin{center}
\includegraphics[height=8cm]{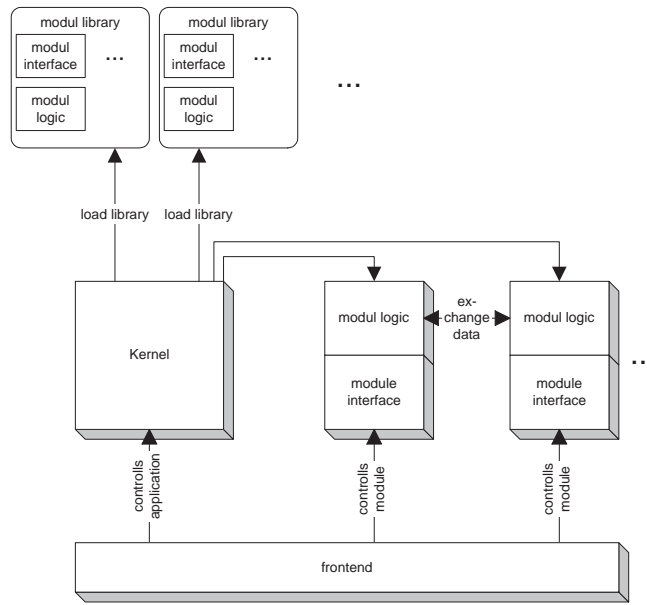}
\end{center}
\caption{Schematic layout of the components of \textbf{openDaVE}.}
\label{opendave-architecture}
\end{figure}

The collaboration of the three parts of \textbf{openDaVE} is shown in figure \ref{opendave-architecture}.

\subsection{Front-ends}
The main aspect of the front-end in view of architecture is its exchangeability. It has access to the application via the
module and kernel interface. These two interfaces are designed to implement a wide variety of front-ends.

\begin{figure}[htc]
\begin{center}
\includegraphics[height=8cm]{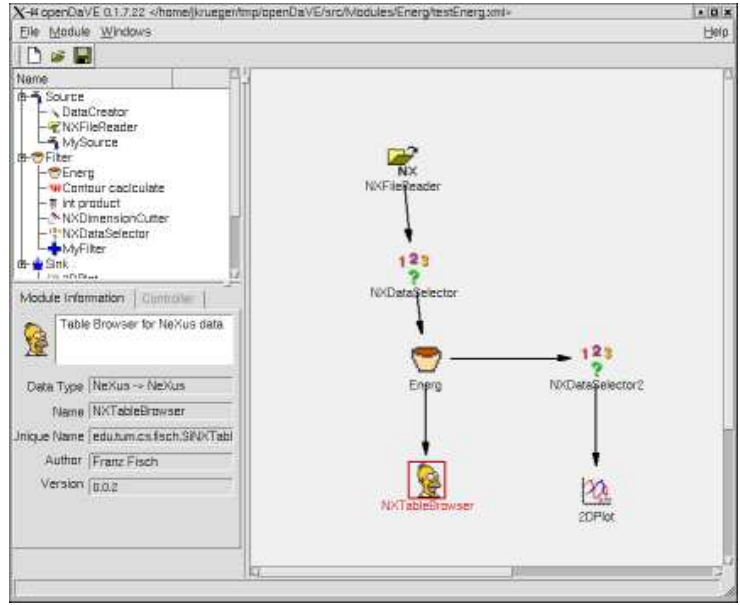}
\end{center}
\caption{Screenshot from a running qdave}
\label{qdave-front-end}
\end{figure}

At this time we have implemented these types of front-ends:
\begin{itemize}
\item Qt based front-end (see figure \ref{qdave-front-end}) \texttt{qdave}
\item Text based front-end \texttt{tdave} 
\item Python bases front-end (under development)
\end{itemize}

\subsection{The Modules}
The actual functionality of the framework is incorporated into the modules. That means an extension of the
functionality of the whole framework is done by developing a new module. 

A module is realised as a shared library being loaded dynamically by the kernel at runtime, on 
request of the user. This library provides a well defined interface to the kernel (the kernel accesses the module logic through
this interface), the module logic (functionality), 
and one or more interfaces to the user which may be used by the front-ends to control the behaviour of the module
logic. So the user may set and get the module parameters.

The creation and deletion of a module instance is done by the kernel. If a module instance is created by the kernel,
exactly one user interface is provided although the library may provide more than one.

\subsubsection{Module functionality}
There exist three types of modules:
\begin{itemize}
\item Sources
\item Filters
\item Sinks
\end{itemize}
The differences between these types are the following: 
\begin{itemize}
\item a source has only output data in the meaning of the \textbf{openDaVE} architecture (for example reading the NeXus data file)
\item a sink has only input data (e.g. writing the data into a NeXus file, display the data)
\item whereas a filter has input as well as output data (for example smoothing a spectrum). 
\end{itemize}
Each module has a certain input and output type which may differ inside one  module. 

\begin{figure}[htc]
\begin{center}
\includegraphics[height=8cm]{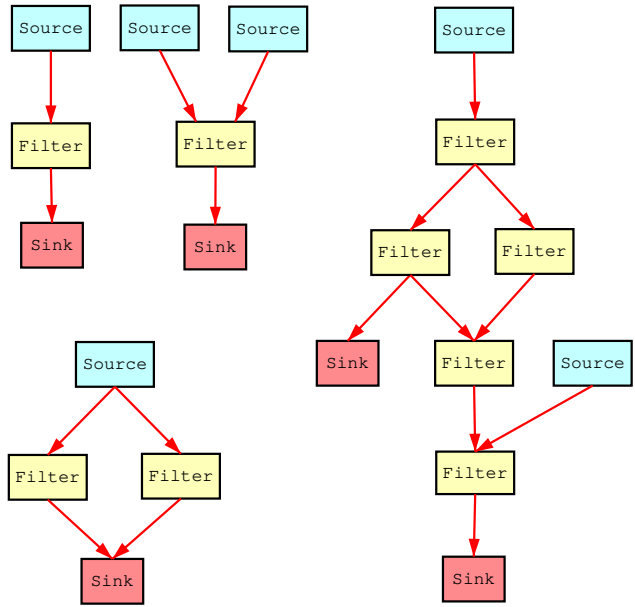}
\end{center}
\caption{Some examples for the data flow in \textbf{openDaVE}}
\label{opendave-dataflow-examples}
\end{figure}

The data in \textbf{openDaVE} may only flow from a source trough one or more filters to a sink. One may consider the data flow as
a chain of modules which has to start with a source (or filter) and has to end with a sink (or filter) (see figure 
\ref{opendave-dataflow-examples}).
The kernel ensures that only modules with the same output and input parameter may be connected.

The functionality of a module is encapsulated in a single function call. This function is always called when the kernel
decides that a module has to recalculate its data. It receives the data from all in-links as parameters. The calculated data
of the module must be returned at the end of the function call. So the data can be delegated by the kernel to all out-links
of the module. The number of in-links can be restricted by the developer of the module.

The data flow may be initialized by a source (a source module creates
new output data, which should be processed by the following module chain) via filters to the 
sink in the so-called real-time mode, or from a sink (the user wants to reprocess the data with changed module parameters)  
via filters to the source. In any case the data flow from source to sink. 

The mode may be set separately for each module, and the modules may be used also in a mixed mode. However, a source initiated data flow
breaks at a module which is not set to the real-time mode.

\subsection{The Kernel}
The kernel may be considered as the heart of \textbf{openDaVE}. Its main task is to manage the modules, meaning that 
it has to create and delete some instances of the modules. Additionally it has to establish and remove the
dataflow between the modules. Let's have a deeper look into the kernel.


Through the front-end instructs the user the
kernel to create or to delete a module instance. Let's consider the creation process. The kernel needs an information about
the type of module to be created, the interface the module has to provide, and the name of the module instance. In its table of 
modules the kernel looks for the type of module which it may create (This table is filled up when the kernel 
loads the different module libraries).
If the type was found the kernel creates a module instance. The module itself tests the existence of the desired interface.
If all searching was succesful the module instance is created and may be used trough the interface by the user.

After creating of the module instances they have to connect among themselves. This connection will be established by the kernel on
request of the user. It checks whether the connection (link) is going from a source/filter to a filter/sink
(see figure \ref{opendave-dataflow-examples}).
Futhermore it checks, whether the output and input data types are the same, and that the maximum number of input links 
to the ending module is not reached.

The deletion of a module requires some works of the kernel. At first it has to check the input and output links to this module. 
If there exist any, it has to stop the dataflow between the modules through them. Afterwards it removes these links before it 
removes the module itself.

\section{Summary}
This paper gives an overview on \textbf{openDaVE}, a modular, open, extensible framework for data analysis and data interpretation.
It is shown how the several parts of the framework collaborate.

\end{document}

%% file: econfmacros.tex
\def\beq{\begin{equation}}
\def\eeq#1{\label{#1}\end{equation}}
\def\eeqn{\end{equation}}

\def\beqa{\begin{eqnarray}}
\def\eeqa#1{\label{#1}\end{eqnarray}}
\def\eeqan{\end{eqnarray}}

\let\bar=\overbar

\def\Dslash{\not{\hbox{\kern-4pt $D$}}}
\def\dslash{\not{\hbox{\kern-2pt $\del$}}}

\def\msb{{\bar{\ssstyle M \kern -1pt S}}}

